\documentclass[runningheads]{llncs}
\usepackage{graphicx}
\usepackage{algorithm}
\usepackage[noend]{algpseudocode}
\usepackage[table,xcdraw]{xcolor}
\usepackage{caption}
\usepackage{hhline}
\usepackage{multirow}
\usepackage{tabularx}
\usepackage{booktabs} 
\usepackage{amsmath}
\usepackage{amssymb}
\usepackage{hyperref}
\hypersetup{
    colorlinks = true,
    linkbordercolor = {white},
    linkcolor = [red]
}

\begin{document}
\title{Whole Slide Images are 2D Point Clouds: Context-Aware Survival Prediction using Patch-based Graph Convolutional Networks}
\titlerunning{Whole Slide Images are 2D Point Clouds}
%
\author{Richard J. Chen, Ming Y. Lu, Muhammad Shaban, Chengkuan Chen, \\Tiffany Y. Chen, Drew F. K. Williamson, Faisal Mahmood}
%
\authorrunning{RJ Chen et al.}
%
\institute{
Department of Pathology, Brigham and Women's Hospital\\
Department of Biomedical Informatics, Harvard Medical School\\
Cancer Data Science Program, Dana-Farber Cancer Institute\\
Cancer Program, Broad Institute of Harvard and MIT\\
\email{richardchen@g.harvard.edu, faisalmahmood@bwh.harvard.edu}}
\maketitle              
\vspace{-5mm}
\begin{abstract}
Cancer prognostication is a challenging task in computational pathology that requires context-aware representations of histology features to adequately infer patient survival. Despite the advancements made in weakly-supervised deep learning, many approaches are not context-aware and are unable to model important morphological feature interactions between cell identities and tissue types that are prognostic for patient survival. In this work, we present Patch-GCN, a context-aware, spatially-resolved patch-based graph convolutional network that hierarchically aggregates instance-level histology features to model local- and global-level topological structures in the tumor microenvironment. We validate Patch-GCN with 4,370 gigapixel WSIs across five different cancer types from the Cancer Genome Atlas (TCGA), and demonstrate that Patch-GCN outperforms all prior weakly-supervised approaches by 3.58-9.46\%. Our code and corresponding models are publicly available at \href{https://github.com/mahmoodlab/Patch-GCN}{https://github.com/mahmoodlab/Patch-GCN}. 
\end{abstract}
%
%
\section{Introduction}

Weakly-supervised deep learning has made remarkable progress in computational pathology in using whole slide images (WSIs) for cancer diagnosis and prognosis \cite{yu2016predicting,campanella2019clinical,courtiol2019deep,wulczyn2020deep,lu2020data,lu2021ai}. Due to the computational complexities in training with WSIs, many weakly-supervised methods have approached WSIs using multiple instance learning (MIL), in which: 1) small image patches from the WSI are extracted as independent instances, and then 2) pooled using a global aggregation operator over the bag of unordered instances. Despite not being context-aware and without needing detailed clinical annotation, many of these MIL-based approaches are able to still solve difficult tasks such as cancer grading and subtyping using only slide-level labels, as the distinction between morphological phenotypes such as tumor vs. non-tumor tissue may only depend on instance-level patch-based features \cite{bandi2018detection,lu2019semi}.

\begin{figure*}[h]
\includegraphics[width=\textwidth]{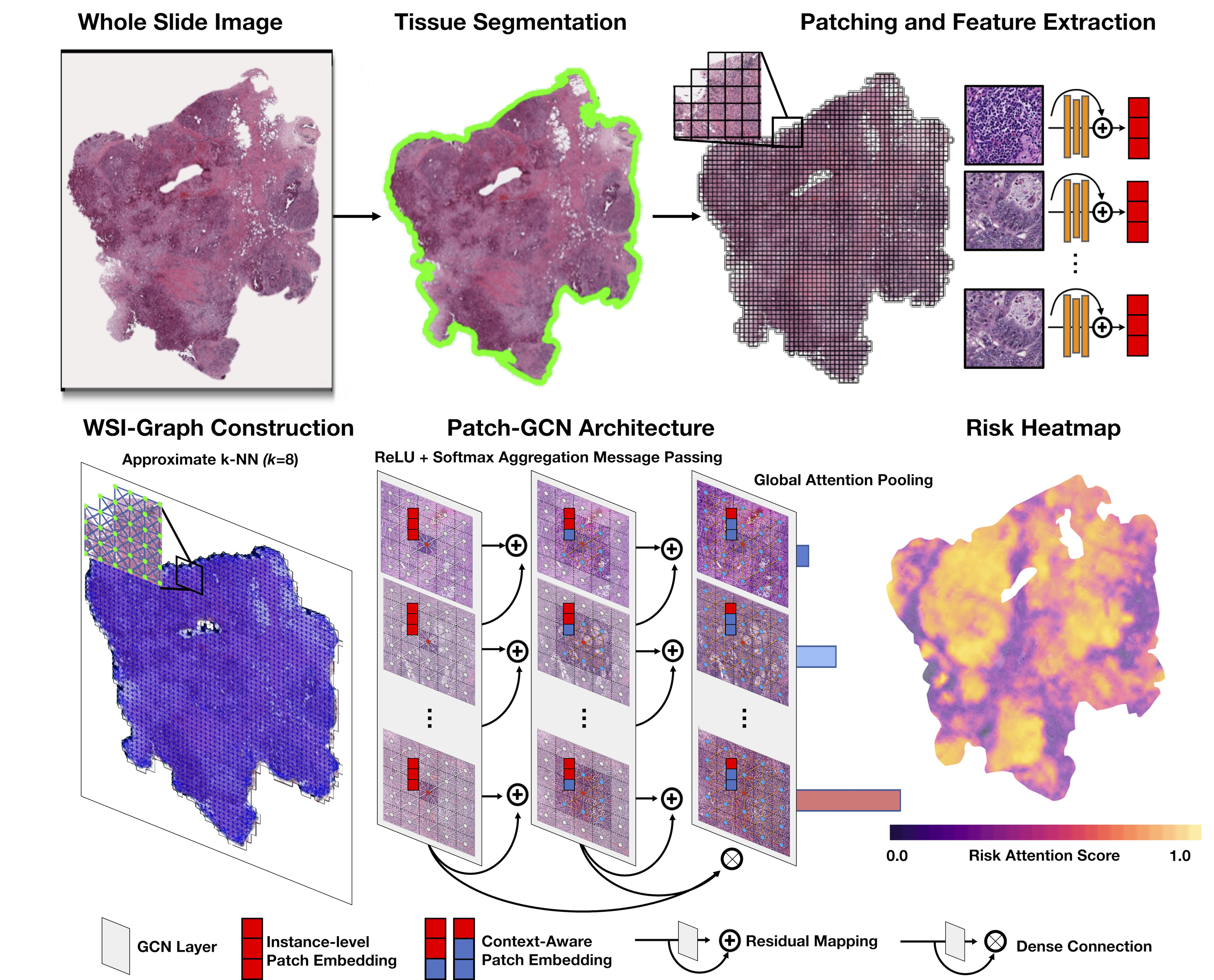}
\caption*{Fig 1: Patch-GCN framework for context-aware survival outcome prediction in WSIs. Non-overlapping $256 \times 256$ patches are patched as used as input into a ResNet-50 CNN to construct the node feature matrix, with edges drawn between adjacent image patches in the WSI. A ReLU + Softmax Message Passing scheme is used to aggregate instance-level embeddings in local neighborhoods, with residual mappings and skip connections used to construct context-aware embeddings, followed by global attention-based pooling.}
\vspace{-5mm}
\end{figure*}

In contrast with cancer grading and subtyping, cancer prognostication is a challenging task that requires considering both instance- and global-level features in the tumor and surrounding tissues for assessing patient risk of mortality \cite{balkwill2012tumor,saltz2018spatial}. In adapting the MIL framework to WSIs, many approaches follow the standard multiple instance (SMI) assumption for solving clinical tasks in computational pathology, \textit{e.g.} - if a bag contains at least one positive instance, it is labeled positive, else negative \cite{zhao2020predicting}. This assumption holds when the clinical task is solving binary instance-level feature discrimination problems such as tumor vs. non-tumor tissue. However, in tasks such as survival outcome prediction in cancer pathology, MIL-based approaches are unable to capture important contextual and hierarchical information that have known prognostic significance in cancer survival \cite{beck2011systematic,abduljabbar2020geospatial,diao2021human,jaume2021quantifying}. For example, though MIL would be able to learn instance-level features that discriminate image patches of lymphocytes and tumor cells, it is unable to distinguish whether those immune cells are tumor-infiltrating lymphocytes (TILs) or from an adjacent inflammatory response, which depends on the lymphocytes' apposition to tumor cells or normal stroma respectively \cite{saltz2018spatial,shaban2019novel,shaban2021digital}.


In this work, we propose a context-aware, spatially-resolved patch-based graph convolutional network (Patch-GCN) for survival prediction in patients with multiple WSIs (Fig. 1). One of the key contributions of our work is that we formulate WSIs as a graph-based data structure in the Euclidean space similar to a point cloud in which: 1) nodes correspond to histology image patches, and 2) edges are connected between adjacent image patches from the true spatial coordinates of the WSI. As a result, message passing in Patch-GCN generalizes the standard convolutional operator in CNNs, in which node features are hierarchically aggregated from local to global structures in the WSI. Compared to other weakly-supervised learning approaches such as MIL, Patch-GCN is context-aware and is able to build hierarchical representations of morphological image patch features in context with their surrounding environment. To robustly validate Patch-GCN, we quantitatively assessed our model on five different cancer datasets from The Cancer Genome Atlas (TCGA) in survival outcome prediction against several state-of-the-art methods in weakly-supervised learning for WSIs, and evaluated the interpretability of Patch-GCN through attention heatmaps in low and high risk patients (Fig. 2). Our code is made available at \href{https://github.com/mahmoodlab/Patch-GCN}{https://github.com/mahmoodlab/Patch-GCN}.

\section{Related Work} 
\subsection{Survival Analysis in WSIs}
In recent years, deep learning methods using CNNs and MIL-based approaches have been proposed for survival analysis in WSIs \cite{lu2018nuclear,mobadersany2018predicting,zhu2017wsisa,wulczyn2020deep,lu2020federated}. Due to the large image sizes of WSIs, many of these methods rely on selective sampling of small image ROIs for tractable training and inference, which are then used matched with patient-level outcome labels. Mobadersany \textit{et al.} \cite{mobadersany2018predicting} proposed one of the first methods for end-to-end training with $1024 \times 1024$ image ROIs using CNNs supervised with the Partial Cox Proportional Hazard loss. Zhu \textit{et al.} \cite{zhu2017wsisa} developed a two-step-based approach for WSI-level survival outcome prediction, in which patches are clustered using K-Means groups using K-Means clustering method then used as inputs into a CNN. Yao \textit{et al.} \cite{yao2020whole} similarly proposed patch-based sampling K-Means clustering to identify morphological phenotypes in WSIs.



\subsection{Graph-based Analysis in Computational Pathology}
In addition to CNNs and MIL-based approaches, GCNs and other graph-based methods have received attention in computational pathology, solving problems such as cancer classification \cite{ali2013cell,anand2020histographs,zhao2020predicting,raju2020graph,ding2020feature,pati2021hierarchical,lu2020capturing}, cancer grading \cite{zhou2019cgc,wang2020weakly,javed2020multiplex}, and survival analysis \cite{li2018graph,chen2020pathomic}. Many of these approaches, however, consider only cell identities as graph nodes, which ignores important prognostic tissue features such as stroma and are confined again to small image regions \cite{beck2011systematic,mahmood2018seg,zhou2019cgc}. In survival analysis, Chen \textit{et al.} \cite{chen2020pathomic} constructed a cell-based graph for small image ROIs followed by spectral convolutions. Alternatively, Li \textit{et al.} \cite{li2018graph} proposed sampling patches in a WSI as nodes, followed by constructing edges between patches via feature similarity on the embedding space and using spectral convolutions. However, we argue that in using this approach for graph construction, GCNs are unable to learn context-aware features as message passing as feature interactions between adjacent image patches are not modeled.

\section{Method} 


\subsection{WSI-Graph Construction} For a given sample, let patient $P$, overall survival time $T$ and censorship status $C$ be a single triplet observation in a dataset $\{P_i, T_i, C_i\}_{i=1}^N$. In addition, let $\{W_{j}\}_{j=1}^K \in P$ be the set of all WSIs for $P$, as there may exist multiple WSIs collected for a single patient. To construct graph $G$ for $P$, we first perform automatic tissue segmentation for all $W_j$ by: 1) transforming a low-downsampled version of $W_j$ into HSV colorspace, and then 2) using Otsu's Binarization on the saturation channel to separate H\&E-stained tissue from the background. Then, non-overlapping $256 \times 256$ instance-level image regions at $20\times$ magnification are patched and used as input for a truncated ResNet-50 model pretrained on ImageNet, which extracts a 1024-dimensional feature vector $h \in \mathbb{R}^{1024}$ via spatial average pooling after the 3rd residual block and is then packed into a node feature matrix $X_{j} \in \mathbb{R}^{m \times 1024}$ for $M_j$ total patches in $W_{j}$. For each patch, we save (x,y)-coordinates from the tissue segmentation, from which we use to build an adjacency matrix $A_{j}$ for each $W_{j}$ via fast approximate k-NN $(k=8)$ that models a $3 \times 3$ image receptive field in CNN convolutions. Finally, we build a subgraph $G_{j} = (X_{j}, A_{j})$, with the patient-level graph across all WSIs constructed as $G = \{G_j\}_{j=1}$ which we denote as a WSI-Graph.

\noindent In comparison to previous graph-based approaches that build neighborhoods using nearest neighbors in the embedding space, our approach is distinct in that graphs are constructed in the Euclidean space. As a result, WSI-Graphs are effectively 2D point clouds (e.g. nodes / points connected to other proximal points in a 2D planar grid), which allows us to leverage spatial convolutions that perform local neighborhood aggregation functions similar to CNNs. In comparison to CNNs, however, Path-GCN is able to tractably perform CNN-like convolution operations on thousands on extracted instance-level image features. 

\subsection{Patch-GCN Architecture} 

\textbf{Message Passing:} For a WSI-Graph $G$ with $M$ instances, we learn a differentiable function $\mathcal{F}_{\text{GCN}}: \mathbb{R}^{M \times d_{in}} \rightarrow \mathbb{R}^{M \times d_{out}}$ parameterized using a GCN that iteratively aggregates and combines node features in their spatial neighborhoods across different hidden layers via message passing. For instance, for the message passing of vertex $v$ (that has node feature $\mathbf{h}_{v}$) with its neighboring vertices $u \in \mathcal{N}(v)$ in hidden layer $G^{(l)}$, we use the graph convolution layer $\mathcal{F}_{\text{GCN}}^{(l)}(G^{(l)}; \phi^{(l)}, \rho^{(l)}, \zeta^{(l)})$ that implement the following functions:
\begin{equation} \label{eq2}
\begin{split}
\textbf{m}_{v}^{(l)}&=\rho^{(l)}\left(\left\{\phi^{(l)}\left(\mathbf{h}_{v}^{(l)}, \mathbf{h}_{u}^{(l)}, \mathbf{h}_{e_{v u}}^{(l)}\right) \rightarrow \mathbf{m}_{v u}^{(l)}: u \in \mathcal{N}(v)\right\}\right) \\ 
\textbf{h}_{v}^{(l+1)}&=\zeta^{(l)}\left(\mathbf{h}_{v}^{(l)}, \mathbf{m}_{v}^{(l)}\right)
\end{split}
\end{equation}

where $\phi^{(l)}$ is a message construction function that calculates a message $\mathbf{m}_{v u}^{(l)}$ between $\mathbf{h}_{v}$ and its neighbor $\mathbf{h}_{u}$ (with edge feature $\mathbf{h}_{e_{v u}}^{(l)}$), $\rho^{(l)}$ is a permutation invariant aggregation function that aggregates all messages passed to $\mathbf{h}_{v}$, and $\zeta^{(l)}$ is an update function that updates the existing node feature at $v$ with the aggregated message $\textbf{h}_{v}^{(l+1)}$. Note that the $\phi^{(l)}, \rho^{(l)}$ in Equation 2 use similar instance-level and bag-level functions in MIL \cite{zaheer2017deep}, in which GCN layers can be considered as performing multiple MIL operations in local graph neighborhoods, with $\zeta^{(l)}$ used as an additional differentiable function for propagating bag-level features across hidden layers in a neural network. In viewing neighborhood aggregation in GCNs has a formulation of MIL with structural neighborhood constraints, we adapt the message passing functions from DeepGCN \cite{li2019deepgcns} which implement $\phi^{(l)}, \rho^{(l)}, \zeta^{(l)}$ as:

\begin{equation} \label{eq3}
\begin{split}
\phi^{(l)}\left(\mathbf{h}_{v}^{(l)}, \mathbf{h}_{u}^{(l)}, \mathbf{h}_{e_{v u}}^{(l)}\right) & = \operatorname{ReLU}\left(\mathbf{h}_{u}^{(l)}+\mathbf{1}\left(\mathbf{h}_{e_{v u}}^{(l)}\right) \cdot \mathbf{h}_{e_{v u}}^{(l)}\right)+\epsilon \rightarrow \mathbf{m}_{v u}^{(l)} \\
\rho^{(l)}\left(\left\{\mathbf{m}_{v u}^{(l)}: \forall u \in \mathcal{N}(v)\right\}\right) &= \sum_{u \in \mathcal{N}(v)} \frac{\exp \left(\beta \mathbf{m}_{v u}^{(l)}\right)}{\sum_{u \in \mathcal{N}(v)} \exp \left(\beta \mathbf{m}_{v u}^{(l)}\right)} \cdot \mathbf{m}_{v u}^{(l)} \rightarrow \textbf{m}_{v}^{(l)} \\
\zeta^{(l)}\left(\mathbf{h}_{v}^{(l)}, \mathbf{m}_{v}^{(l)}\right) &= \mathbf{M L P}\left(\mathbf{h}_{v}^{(l)}+\mathbf{m}_{v}^{(l)}\right) \rightarrow \textbf{h}_{v}^{(l+1)}
\end{split}
\end{equation}

\noindent in which $\phi^{(l)}$ is the additively combines node and edge features followed by $\operatorname{ReLU}$ activation, $\rho^{(l)}$ is a Softmax Aggregation scheme similar to Ilse \textit{et al.} \cite{ilse2018attention} that computes an attention weight $a_{vu}^{(l)}$ that weights how much $\mathbf{m}_{v u}^{(l)}$ should contribute to the aggregated message $\mathbf{m}_{v}^{(l)}$, and $\zeta^{(l)}$ additive combines the current node feature and aggregated message followed by a multilayer perceptron. Additionally, $\mathbf{1}(\cdot)$ is an indicator function when an edge feature $\mathbf{h}_{e_{v u}}^{(l)}$ exists, $\epsilon$ is a positive constant for numerical stability (set to $10^{-7}$), and $\beta$ is a hyperparameter for the inverse temperature in Softmax (set to $1$). We argue that $\rho^{(l)}$ can be viewed as a formulation of attention pooling operation in Ilse \textit{et al.} \cite{ilse2018attention} with structural neighborhood constraints, in which attention pooling of instance-level features is performed in local graph neighborhoods instead of across the entire bag.

\noindent\textbf{Learning Hierarchical Features:} To learn global-level morphological features in WSIs, following \cite{li2019deepgcns}, we make $\mathcal{F}_{\text{GCN}}^{(l)}$ a residual mapping and stack multiple layers of $\mathcal{F}_{\text{GCN}}^{(l)}$ where the output of $\mathcal{F}_{\text{GCN}}^{(l)}$ additively combines with its input.
\begin{equation}
    G^{(l+1)} = \mathcal{F}_{\text{GCN}}^{(l)}(G^{(l)}; \phi^{(l)}, \rho^{(l)}, \zeta^{(l)})+G^{(l)}
\end{equation}

\begin{figure}[h]
\includegraphics[width=\textwidth]{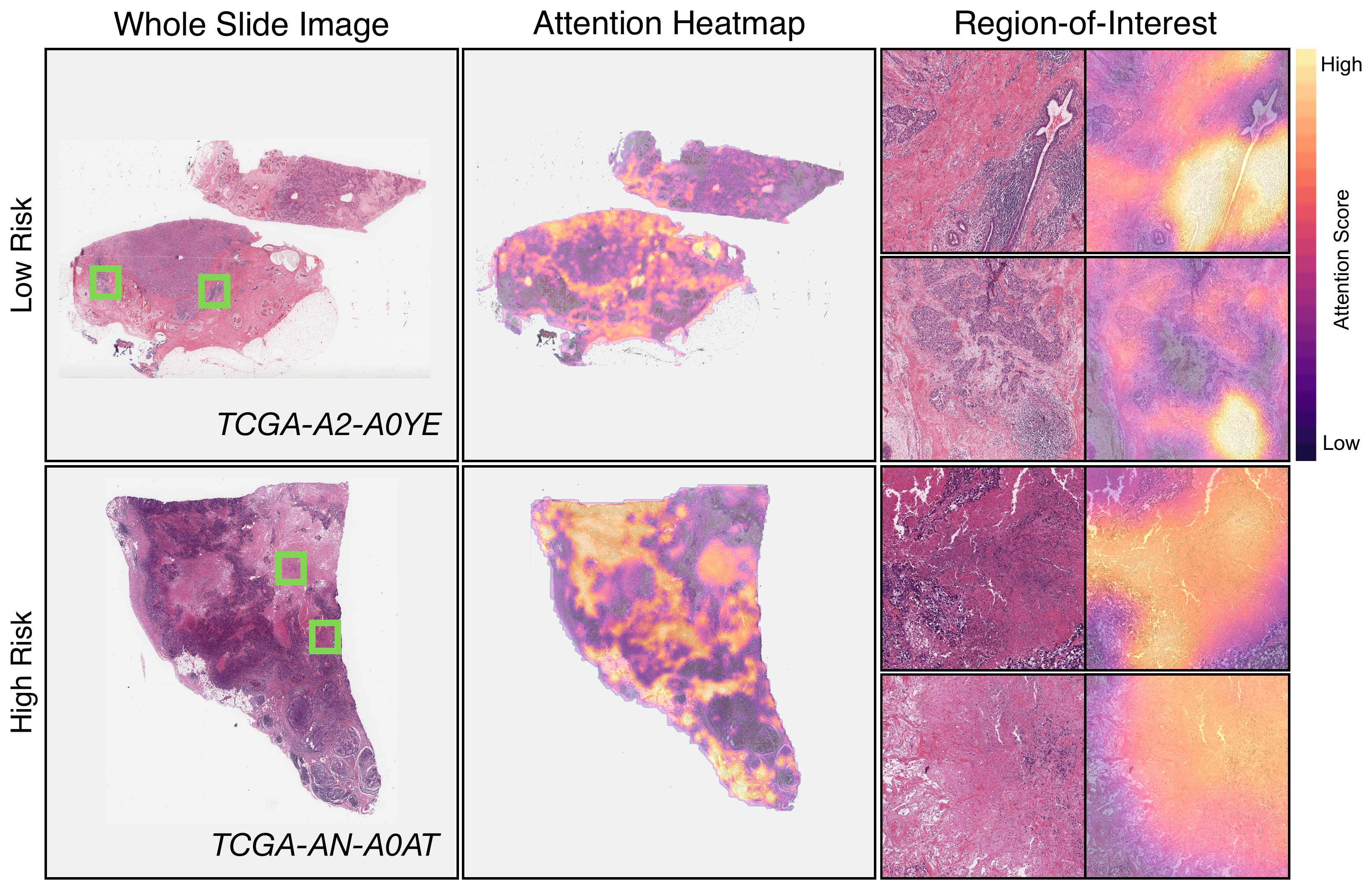}
\caption*{Fig. 2: Patch-GCN interpretability in BRCA survival prediction. In low risk patients, high attention regions corresponded to aggregates of lymphocytes near tumor cells, whereas in high risk patients, high attention regions corresponded to areas of tumor-associated stroma and necrosis.}
\vspace{-5mm}
\end{figure}

\noindent We implement the spatial neighborhood aggregation backbone of Patch-GCN using $L=4$ graph convolutional layers. As a result, each patch-based histology image feature aggregates features from other nodes in a 4-hop neighborhood, which results in an effective image receptive field size of $2302 \times 2302$ for $256 \times 256$ patches connected to its 8 nearest neighbors (Fig. 3, Supplementary Material). Furthermore, we also implement dense connections from the output of every GCN Layer to the last hidden layer of $\mathcal{F}_{\text{GCN}}$, so that the representation of each histology patch would be an amalgamation of its instance-level embedding and its learned surrounding context, written as $\mathbf{H}^{(L)} = [X^{(1)}, \dots ,X^{(L)}]$.

\noindent\textbf{Global Neighborhood Aggregation \& Supervision:} From the penultimate node feature matrix $\mathbf{H}^{(L)}$, following \cite{ilse2018attention}, we learn a global attention-based pooling layer $\mathcal{F}_{\text{AttnMIL}}(\mathbf{H}^{(L)}; \phi^{(L)}, \rho^{(L)})$ that adaptively computes a weighted sum of all node features in the graph, which generalizes aggregation function in Equation 2 to function on all nodes in the graph, in which the node feature matrix for the last hidden layer $\mathbf{H}^{(L)} \in \mathbb{R}^{m \times d_{\text{out}}}$ is pooled to a WSI-level embedding $\mathbf{h}_m^{(L)} \in \mathbb{R}^{1 \times d_{\text{out}}}$, which is subsequently supervised using the cross entropy-based Cox proportional loss function following \cite{zadeh2020bias} for survival analysis.

\noindent\textbf{Implementation Details:} To train Patch-GCN, we used Adam optimization with a default learning rate of $2 \times 10^{-4}$, weight decay of $1 \times 10^{-5}$, using a ResNet-50 CNN backbone pretrained on ImageNet, and trained for 20 epochs. To train with large graphs, we used $4$ NVIDIA 2080 Ti GPUs with a batch size of 1 with 32 steps for gradient accumulation.

\section{Experimental Setup}

For this study, we used 4,370 diagnostic gigapixel WSIs across five different cancer types from The Cancer Genome Atlas: Bladder Urothelial Carcinoma (BLCA) $(n=437)$, Breast Invasive Carcinoma (BRCA) $(n=1022)$, Glioblastoma \& Lower Grade Glioma (GBMLGG) $(n=1011)$, Lung Adenocarcinoma (LUAD) $(n=515)$, and Uterine Corpus Endometrial Carcinoma (UCEC) $(n=538)$. Our selection criterion in choosing these cancer types for training and evaluation were defined by: 1) dataset size, and 2) balanced distribution of uncensored-to-censored patients. On average, each WSI contained approximately 13487 $256 \times 256$ image patches at $20\times$ magnification, with some patients having graph sizes as large as $100000$ instances.

\noindent To evaluate Patch-GCN, we trained our proposed model using 5-fold cross-validation for each cancer type, in which each dataset was split into 5 80/20 partitions for training and validation. The cross-validated concordance index (c-Index) across the validation splits was used to measure the predictive performance in correctly ranking the survival times of each patient. As qualitative assessment, we used Kaplan-Meier curves to visualize the quality of patient stratification in stratifying low and high risk patients as two different survival distributions, as well as attention-based heatmaps using the weights computed by $\mathcal{F}_{\text{AttnMIL}}$ (Fig. 2, 4). In addition, we compared Patch-GCN against several other weakly-supervised deep learning approaches for processing in WSIs in computational pathology. As a fair comparison, we used the same survival loss function, ResNet-50 feature embeddings, and training hyperparameters in Patch-GCN.


\section{Results \& Discussion} 
\subsection{Quantitative Results}
In comparing our approach to other weakly-supervised learning methods for WSIs in computational pathology, Patch-GCN outperforms all prior approaches on 4 out of 5 cancer types in head-to-head comparisons, achieving an overall c-Index of $\textbf{0.636}$ (Table 1). For cancer types such as GBMLGG which has known intertumoral and intratumoral heterogeneity, Patch-GCN achieves a c-Index of \textbf{0.824} using WSIs and shows patient stratification into distinct survival groups (Fig. 4, Supplementary Material), which empirically suggests that Patch-GCN is able to learn context-aware features via hierarchical feature aggregation in local spatial neighborhoods. In comparing Patch-GCN to permutation-invariant / MIL-based approaches, we observe that Patch-GCN improves over all methods on all 5 cancer types (9.46\% performance increase over DeepAttnMISL and 3.58\% performance increase over Attention MIL), which further suggests that context matters in survival outcome prediction in WSIs. In comparison to DeepGraphConv which samples random patch features from WSIs as nodes and connects these nodes on the embedding space, Patch-GCN improves on all cancer types except UCEC (2.58\% performance increase), which suggests the importance of building graphs via adjacent patches rather than feature similarity in the embedding space. Though DeepGraphConv has higher c-Index on UCEC, we note that in comparison to other cancer types, cancer prognosis in UCEC correlates with global-level morphological determinants such as tumor size and depth of tumor invasion in the myometrium, rather than cell-to-cell mediated interactions between tumor cells and other cell types. BLCA is a similar cancer type to UCEC that also depends on the depth of invasion into the bladder wall, but because the bladder wall is thinner than the myometrium, the invasion may be adequately captured via a limited receptive field, hence better Patch-GCN performance on that cancer type.

\begin{table*}[h]
\centering
\begin{tabular*}{\textwidth}{l@{\extracolsep{\fill}}ccc}
\toprule
Models &             BLCA &             BRCA &           GBMLGG \\
\midrule
MIL (Deep Sets) \cite{zaheer2017deep}     &  0.500 $\pm$ 0.000 &  0.500 $\pm$ 0.000 &  0.498 $\pm$ 0.014 \\
Attention MIL \cite{ilse2018attention} &  0.536 $\pm$ 0.038 &  0.564 $\pm$ 0.050 &  0.787 $\pm$ 0.028 \\
DeepAttnMISL \cite{yao2020whole}  &  0.504 $\pm$ 0.042 &  0.524 $\pm$ 0.043 &  0.734 $\pm$ 0.029 \\
DeepGraphConv \cite{li2018graph} &  0.499 $\pm$ 0.057 &  0.574 $\pm$ 0.044 &  0.816 $\pm$ 0.025 \\ \vspace{1mm}
Patch-GCN (Ours)     &  \textbf{0.560 $\pm$ 0.034} &  \textbf{0.580 $\pm$ 0.025} &  \textbf{0.824 $\pm$ 0.024} \\ 
{} &            LUAD &             UCEC & Overall \\
\midrule
MIL (Deep Sets) \cite{zaheer2017deep}     &  0.496 $\pm$ 0.008 &  0.500 $\pm$ 0.000 &   0.499 \\
Attention MIL \cite{ilse2018attention} &  0.559 $\pm$ 0.060 &  0.625 $\pm$ 0.057 &   0.614 \\
DeepAttnMISL \cite{yao2020whole}  &  0.548 $\pm$ 0.050 &  0.597 $\pm$ 0.059 &   0.581 \\
DeepGraphConv \cite{li2018graph} &  0.552 $\pm$ 0.058 &  \textbf{0.659 $\pm$ 0.056} &   0.620 \\
Patch-GCN (Ours)     &  \textbf{0.585 $\pm$ 0.012} &  0.629 $\pm$ 0.052 &   \textbf{0.636} \\
\bottomrule
\end{tabular*}
\caption{c-Index performance comparisons of Patch-GCN against prior state-of-the-art weakly-supervised approaches on 5 cancer types in the TCGA.}
\vspace{-6mm}
\end{table*}

\vspace{-2mm}
\subsection{Attention Visualization}
To understand how Patch-GCN uses morphological features to predict risk, we visualized heatmaps using the attention weights from the attention pooling layer and utilized two trained pathologists to assess high-attention image regions. Across all cancers, we observed that in high risk patients, the network assigned high attention to necrosis, dense tumor aggregates, and regions of desmoplastic stroma containing tumor infiltrates, which are indicative of tumor invasion and proliferation (Fig. 2). In low risk patients, we observe that lymphocyte aggregates and normal stroma were frequently assigned high attention, which corroborates with the prognostic significance of stroma \cite{beck2011systematic}. Fig. 2 shows exemplar low and high risk cases in BRCA, with lymphocytes adjacent to tumor cells and infiltrating normal stroma given high attention in low risk patients, while necrosis and desmoplastic stroma were given high attention in high risk patients.

\subsection{Conclusion}
Despite the progress made in weakly-supervised deep learning in computational pathology, many current approaches are not context-aware in modeling important local- and global-level morphological features in the tumor microenvironment. In this work, we present Patch-GCN, a context-aware, attention-based graph convolutional network for survival analysis using WSIs. In comparing Patch-GCN to permutation-invariant network architectures that learn only instance-level morphological features, we observe that Patch-GCN outperforms all prior approaches on 5 cancer types in the TCGA. Moreover, we demonstrate the improvement in connecting nodes via adjacent image patches, which allows node aggregation in GCNs to learn such coarse-grained to fine-grained topological structures in the tumor microenivronment. Our approach is adaptable to any weakly-supervised learning task in computational pathology that uses slide-level or patient-level labels, and contributes towards a more holistic view of representation learning in the tumor microenvironment.

\section{Acknowledgements}
\noindent\textbf{Funding:} This work was supported in part by internal funds from BWH Pathology, Google Cloud Research Grant, Nvidia GPU Grant Program, and NIGMS R35GM138216 (F.M.). R.J.C. was additionally supported by the NSF Graduate Fellowship. The content is solely the responsibility of the authors and does not reflect the official views of the National Institutes of Health, National Institute of General Medical Sciences or the National Science Foundation. \\
\vspace{-5mm}
%
%
%
%
\newpage
\bibliographystyle{splncs03_unsrt}
\bibliography{paper2410.bib}

\vspace{-10mm}
\begin{figure*}[h]
\section*{Appendix}
\vspace{3mm}
\includegraphics[width=\textwidth]{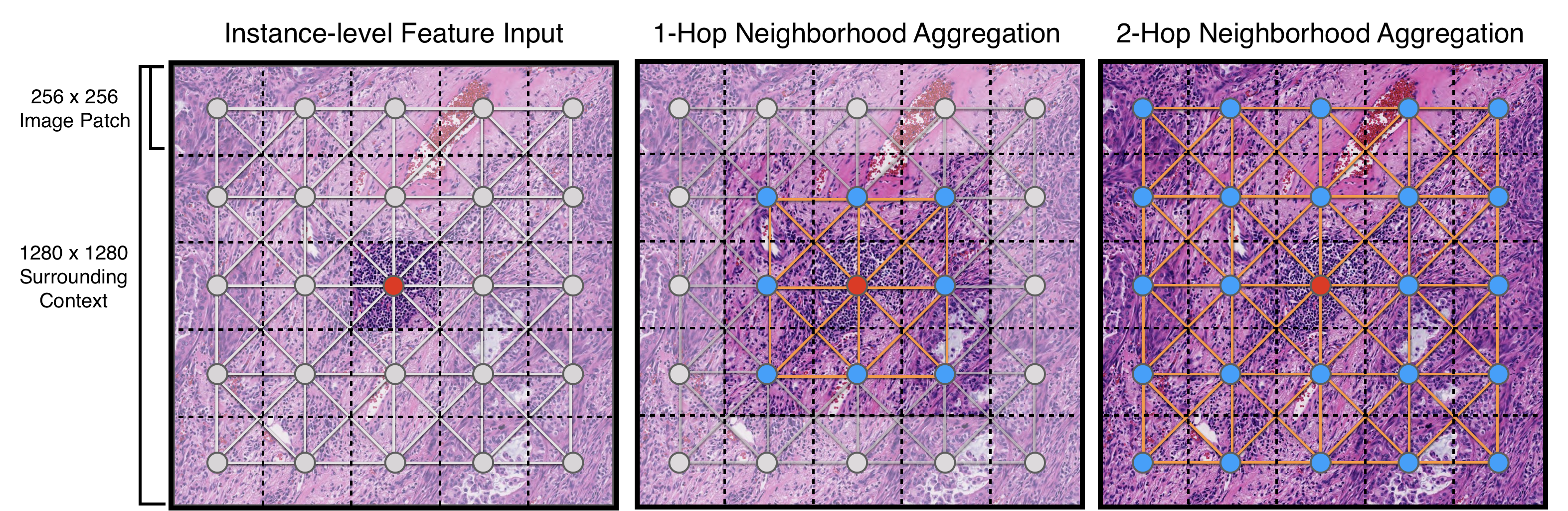}
\caption*{Fig. 3: Feature Aggregation in Patch-GCN. Red shows the histology image patch-of-interest, blue and orange shows active nodes and edges respectively, and grey / low-opacity represent inactive nodes. From instance-level feature input (left), we extract features from the highlighted $256 \times 256$ histology image patch $x_m$ ($m$th instance in the WSI) as a $\textbf{h}_m^{(0)} \in \mathbb{R}^{1 \times 1024}$ (containing only lymphocytes), which we use to initialize the node feature matrix in $\mathcal{F}_{\text{GCN}}$. The 1-hop neighborhood for $\textbf{h}_m^{(0)}$ is a $3 \times 3$ receptive field in the graph neighborhood ($768 \times 768$ image context around $x_m$), which aggregates into $\textbf{h}_m^{(1)}$ (middle). As we perform message passing, the context around $x_m$ expands to aggregate more distal morphological features (tumor cells, stroma, thrombosis) (right).}
\vspace{3mm}
\includegraphics[width=\textwidth]{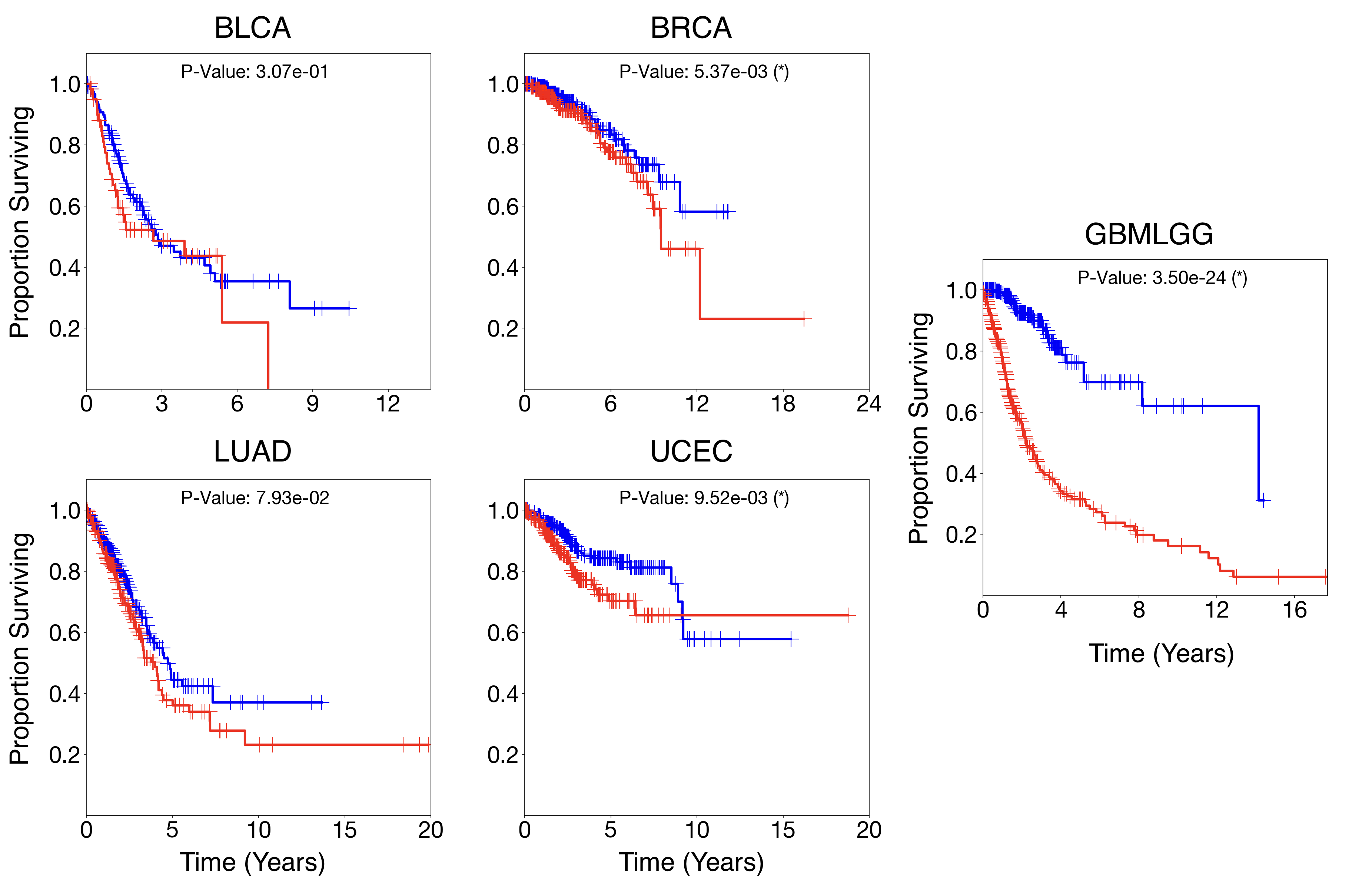}
\caption*{Fig. 4: Kaplan-Meier Analysis of Patch-GCN. Out-of-sample risk predictions from each validation fold were pooled and then plotted against their survival time. For significance testing, we use the logrank test to measure the statistical difference between two survival distributions that correspond to low risk (blue) and high risk (red) patients (P-Value $<$ 0.05).}
\end{figure*}

\end{document}